\documentclass[conference]{IEEEtran}
\IEEEoverridecommandlockouts
\usepackage{atbegshi}
\AtBeginDocument{\AtBeginShipoutNext{\AtBeginShipoutDiscard}}
\usepackage{caption}
\usepackage{booktabs}
\usepackage{float}
\usepackage{flushend}
\usepackage{multirow} 
\usepackage{threeparttable} 
\usepackage[nolist]{acronym}
\usepackage{cite}
\usepackage{amsmath,amssymb,amsfonts}
\usepackage{algorithmic}
\usepackage{graphicx}
\usepackage{textcomp}
\usepackage{xcolor}
\def\BibTeX{{\rm B\kern-.05em{\sc i\kern-.025em b}\kern-.08em
    T\kern-.1667em\lower.7ex\hbox{E}\kern-.125emX}}
\begin{document}{

\title{ONNX-to-Hardware Design Flow for the Generation of Adaptive Neural-Network Accelerators on FPGAs}*\\

}

\author{\IEEEauthorblockN{1\textsuperscript{st} Federico Manca}
\IEEEauthorblockA{\textit{Dipartimento di Ingegneria Elettrica ed Elettronica} \\
\textit{Università degli Studi di Cagliari}\\
Cagliari, Italy \\
f.manca58@studenti.unica.it}
\and
\IEEEauthorblockN{2\textsuperscript{nd} Francesco Ratto}
\IEEEauthorblockA{\textit{Dipartimento di Ingegneria Elettrica ed Elettronica} \\
\textit{Università degli Studi di Cagliari}\\
Cagliari, Italy \\
0000-0001-5756-5879}
}

\maketitle



\begin{acronym}
\acro{ac}[AC]{Approximate Computing}
\acro{aes}[AES]{Advanced Encryption Standard}
\acro{api}[API]{Application Programming Interface}
\acrodefplural{api}[APIs]{Application Programming Interfaces }
\acro{asic}[ASIC]{Application Specific Integrated Circuit}

\acro{bnn}[BNN]{Binary Neural Network}
\acrodefplural{bnn}[BNNs]{Binary Neural Networks}

\acro{cgr}[CGR]{Coarse-Grain Reconfigurable}
\acro{cpg}[CPG]{Co-Processor Generator}
\acro{cps}[CPS]{Cyber-Physical System}
\acrodefplural{cps}[CPSs]{Cyber-Physical Systems}
\acro{cpu}[CPU]{Central Processing Unit}
\acrodefplural{cpu}[CPUs]{Central Processing Units}

\acro{dag}[DAG]{Directed Acyclic Graph}
\acro{dnn}[DNN]{Deep Neural Network}
\acro{dpn}[DPN]{Dataflow Process Network}
\acrodefplural{dpn}[DPNs]{Dataflow Process Networks}
\acro{dse}[DSE]{Design Space Exploration}
\acro{dsp}[DSP]{Digital Signal Processing}


\acro{fft}[FFT]{Fast Fourier Transform}
\acro{fifo}[FIFO]{First-In First-Out queue}
\acrodefplural{fifo}[FIFOs]{First-In First-Out queues}
\acro{fpga}[FPGA]{Field Programmable Gate Array}
\acrodefplural{fpga}[FPGAs]{Field Programmable Gate Arrays}

\acro{gpu}[GPU]{Graphics Processing Unit}


\acro{hevc}[HEVC]{High Efficiency Video Coding}
\acro{hls}[HLS]{High Level Synthesis}
\acro{hwpu}[HWPU]{HW Processing Unit}
\acrodefplural{hwpu}[HWPUs]{HW Processing Unit (HWPU)s}

\acro{ip}[IP]{Intellectual Property}
\acrodefplural{ip}[IPs]{Intellectual Properties}


\acro{lut}[LUT]{Look-Up Table}
\acrodefplural{lut}[LUTs]{Look-Up Tables}

\acro{mdc}[MDC]{Multi-Dataflow Composer}
\acro{mdg}[MDG]{Multi-Dataflow Generator}
\acro{mlp}[MLP]{Multi Layer Perceptron}
\acro{mcdma}[MCDMA]{Multi-channel DMA}
\acro{moa}[MoA]{Model of Architecture}
\acrodefplural{moa}[MoAs]{Models of Architecture}
\acro{moc}[MoC]{Model of Computation}
\acrodefplural{moc}[MoCs]{Models of Computation}
\acro{mpsoc}[MPSoC]{Multi-Processor \ac{soc}}
\acrodefplural{mpsoc}[MPSoCs]{Multi-Processor System on Chip}

\acro{nn}[NN]{Neural Network}

\acro{os}[OS]{Operating System}

\acro{pc}[PC]{Platform Composer}
\acro{pe}[PE]{Processing Element}
\acro{pisdf}[PiSDF]{Parameterized and Interfaced Synchronous DataFlow}
\acro{pdf}[PDF]{Parameterized DataFlow}
\acro{pl}[PL]{Programmable Logic}
\acro{ps}[PS]{Processing System}
\acro{pmc}[PMC]{Performance Monitoring Counter}
\acro{psdf}[PSDF]{Parameterized Synchronous DataFlow}

\acro{qsoc}[QSoC]{Quantized \ac{soc}}

\acro{qnn}[QNN]{Quantized Neural Network}

\acro{rtl}[RTL]{Register Transfer Level}

\acro{sg}[SG]{Scatter-Gather}
\acro{sdf}[SDF]{Synchronous DataFlow}
\acro{soc}[SoC]{System on a Chip}
\acro{smmu}[SMMU]{System Memory Management Unit}

\acro{til}[TIL]{Template Interface Layer}

\acro{uav}[UAV]{Unmanned Aerial Vehicle}
\acro{ugv}[UGV]{Unmanned Ground Vehicle}





\end{acronym}

\begin{abstract}
\acp{nn} provide a solid and reliable way of executing different types of applications, ranging from speech recognition to medical diagnosis, speeding up onerous and long workloads. 
The challenges involved in their implementation at the edge include providing diversity, flexibility, and sustainability. That implies, for instance,  supporting evolving applications and algorithms energy-efficiently.
Using hardware or software accelerators can deliver fast and efficient computation of the \acp{nn}, while flexibility can be exploited to support long-term adaptivity.
Nonetheless, handcrafting a \ac{nn} for a specific device, despite the possibility of leading to an optimal solution, takes time and experience, and that's why frameworks for hardware accelerators are being developed. 
This work-in-progress study focuses on exploring the possibility of combining the toolchain proposed by Ratto et al.~\cite{ratto2023}, which has the distinctive ability to favor adaptivity, with \ac{ac}. The goal will be to allow lightweight adaptable \ac{nn}  inference on FPGAs at the edge. Before that, the work presents a detailed review of established frameworks that adopt a similar streaming architecture for future comparison.

\end{abstract}


\section{Introduction}
\ac{cps} integrate \textit{``computation with physical processes whose behavior is deﬁned by both the computational (digital and other forms) and the physical parts of the system''}\footnote{https://csrc.nist.gov/glossary/term/cyber physical systems}. They are characterized by a considerable information exchange with the environment and by dynamic and reactive behaviors with respect to environmental changes.
In modern systems, \ac{cps} or not, decisions making can be brought directly at the edge on small embedded platforms exploiting the capabilities of the \acp{nn}.
This calls for real-time, low-energy execution, which can be achieved by leveraging different resources on the same platforms, making heterogeneous computing fundamental.

\acp{fpga} devices can guarantee hardware acceleration, execution flexibility, and energy efficiency, as well as heterogeneity, and that is why this type of device is a valuable option for \acp{nn} inference at the edge~\cite{guo2019}. Indeed, their integration on heterogenous \acp{mpsoc} opened up a wide range of possibilities exploiting the \ac{fpga} potentials along with the CPU capabilities of managing control flow and communication.

While many solutions exist to deploy AI models on these platforms, they often lack full support for advanced features, such as flexibility. This paper focuses on putting the basis to achieve future runtime adaptivity, which is key in our opinion for addressing the dynamic and reactive nature of \ac{cps}. Designing and deploying reconfigurable accelerators with these functionalities still needs to be investigated, requiring an in-depth knowledge of the underlying hardware and hand-tailored solutions. This belief is at the basis of this study that, starting from a state-of-the-art framework \cite{ratto2023}, intends to ultimately seek automated strategies to deploy lightweight, flexible, and adaptive \ac{nn} accelerators for \ac{cps}, achieving different Pareto optimal working points that could be merged into an adaptive accelerator and exploited at runtime to serve variable and evolvable applications.

\section{BACKGROUND}
Several software libraries and frameworks have been developed to facilitate the development and high-performance execution of CNNs. Tools such as Caffe\footnote{https://caffe.berkeleyvision.org/}, CoreML\footnote{https://developer.apple.com/documentation/coreml}, PyTorch\footnote{https://pytorch.org/}, Theano\footnote{https://github.com/Theano} and TensorFlow\footnote{https://www.tensorflow.org/} aim to increase the productivity of CNN developers by providing high-level APIs to simplifying data pipeline development.
Such environments are currently flanked \ac{hls} that are used to generate FPGA-based hardware designs from a high level of abstraction, to fasten porting of complex algorithms at the edge. Examples of \ac{hls} environments are AMD’s Vitis HLS, Intel FPGA OpenCL SDK, Maxeler’s MaxCompiler \cite{maxcompiler}, and LegUp \cite{Legup}. They employ commonly used programming languages such as C, C++, OpenCL, and Java, to fill the gap between software-defined applications and their hardware implementation. 

To execute \ac{nn} at the edge, three main types of architectures can be found in literature~\cite{venieris2018}: the Single Computational Engine architecture, based on a single computation engine, typically in the form of a systolic array of processing elements or a matrix multiplication unit, that executes the CNN layers sequentially~\cite{systolic};
Vector Processor architecture, with instructions specific for accelerating operations related to convolutions~\cite{vector_processor}; 
the Streaming architecture consists of one distinct hardware block for each layer of the target CNN, where each block is optimized separately~\cite{aarestad2021,fraser2017}. In our studies, we focus mainly on the latter.

\subsection{Streaming Architectures} \label{streaming-archs}
In our previous work \cite{ratto2023}, the dataflow model was the best-suited one to support runtime adaptivity and enhancing parallelism. The resulting hardware is a streaming architecture that uses on-chip memory, guaranteeing low-latency and low-energy computing.
Solutions that exploit a similar streaming architecture are FINN \cite{fraser2017}, an experimental framework from AMD Research Labs based on Theano; HLS4ML \cite{aarestad2021}, an open-source software designed to facilitate the deployment of machine learning 
models on FPGAs, targeting low-latency and low-power edge applications.

These two solutions are described in Sections \ref{finn} and \ref{hls4ml} respectively, and their performance is compared in Table~\ref{table::comparison}.

\subsubsection{FINN} \label{finn}
FINN is a framework for building scalable and fast \ac{nn}, with a focus on the support of \ac{qnn} inference on
FPGAs. A given model, trained through Brevitas, is compiled 
by the FINN compiler, producing a synthesizable C++ description of a heterogeneous streaming architecture. All
\ac{qnn} parameters are kept stored in the on-chip memory, which greatly reduces the power consumed and simplifies the design. The computing
engines communicate via the on-chip data stream. Avoiding the ``one-size-fits-all'', an ad-hoc topology is built for the network. The resulting accelerator is deployed on the target board using the AMD Pynq framework. 
Two works adopting the FINN framework have been analyzed and their results are summarized in Table~\ref{table::comparison}. 

\subsubsection{HLS4ML}\label{hls4ml}
The main operation of the HLS4ML library is to translate the model of the network into an HLS Project. The focus in \cite{diguglielmo2020} was centered on reducing the computational complexity and resource usage on a fully connected network for MNIST dataset classification: the data is fed to a multi-layer perceptron with an input layer of 784 nodes, three hidden layers with 128 nodes each, and an output layer with 10 nodes. 
The work exploits the potential of Pruning and Quantization-Aware Training to drastically reduce the model size with limited impact on its accuracy. 

To the best of our knowledge, neither FINN nor HLS4ML, despite targeting \ac{fpga}-based streaming architecture and supporting \ac{ac} features such as pruning and quantization, ever proposed a reconfigurable solution for runtime adaptive environments.

\begin{table*}[ht]

\renewcommand\arraystretch{1} 
	\centering
 \caption{Performance overview of FINN and HLS4ML under different testing set-ups.}
 \begin{threeparttable}
\begin{tabular}{l l c c c l c c c c c c}
\toprule
\multirow{2}{*}{Framework} &\multirow{2}{*}{Dataset}        &FC         &CONV       &Datatype       & Target    &LUTS       &DSP        & Latency       & Throughput    & Power  & Accuracy  \\ 
 &                          & [\#]    & [\#]    & [\# bits]    & board     & [\#]    & [\#]     & [us]          & [FPS]      &[W]   &[\%]   \\ 

\midrule
FINN \cite{umuroglu2017}       &CIFAR-10\tnote{*} & 2           & 6             & 2                 & Zynq7000 & 46253   & -      & 283    & 21.9k   & 15.3          & 80.1 \\ 
FINN \cite{umuroglu2017}      &SVHN\tnote{*} & 2           & 6             & 2                 & Zynq7000 & 46253   & -      & 283    & 21.9k   & 15.3          & 94.3  \\ 
FINN \cite{fraser2017}      & CIFAR-10        & 2           & 6             & 2                 & UltraScale & 392947  & -      & 671       & 12k     & \textless 41  & 88.3                           \\ 
HLS4ML \cite{diguglielmo2020}     & SVHN            & 3           & 3             & 7                 & UltraScale+    & 38795   & 72     & 1035  & -          & -              & 95                             \\ 
HLS4ML \cite{aarestad2021}    & MNIST           & 3           & 0             & 16                & Ultrascale+    & 366494  & 11     & 200    & -          & -              & 96                             \\ \bottomrule
\end{tabular}
     \begin{tablenotes}
    \item[*] The two datasets are cropped to have the same image size.
    \end{tablenotes}
 \end{threeparttable}
\label{table::comparison}
\end{table*}

\subsection{Approximate Computing}\label{ss:ac}
\ac{ac} has been established as a new design paradigm for energy-efficient circuits, exploiting the inherent ability of a large number of applications to produce results of acceptable quality, despite some approximations in their computations. Leveraging this
property, \ac{ac} approximates the hardware execution of the error-resilient computations in a manner that favors performance and energy. Moreover, \acp{nn} have demonstrated strong resilience to errors and can take great advantage of AC \cite{mittal2016}. In particular, hardware \ac{nn} approximation can be classified into three wide categories: Computation Reduction, Approximate Arithmetic Units, and Precision Scaling~\cite{ac_survey}.
\paragraph{Computational Reduction}
The Computation Reduction approximation category aims at systematically
avoiding, at the hardware level, the execution of some computations, significantly decreasing the executed workload. An example of this is pruning: biases, weights, or entire neurons can be evicted to lighten the workload~\cite{comp_reduction}.
\paragraph{Approximate Units}
With Approximate Units, it is improved the energy consumption and latency of DNN accelerators by employing approximate
circuits that replace more accurate units, like the Multiply-and-Accumulate (MAC) one~\cite{approx_unit}.
\paragraph{Precision Scaling}
The most used Precision Scaling practice is quantization:  quantized hardware implementations feature reduced bit-width dataflow and arithmetic units 
attaining very high energy, latency, and bandwidth gains compared to 32-bit floating-point
implementations. Rather than executing all
the required mathematical operations with ordinary 32-bit or 16-bit floating point, quantization allows to exploit smaller integer operations instead \cite{precision_scaling}. 
For this purpose, AMD provides an arbitrary precision data types library for use in Vivado HLS designs, which allows the specification of any number of bits for data types beyond what the standard C++ data types provide. The library also supports customizable fixed-point data types \cite{ap_fixed}.

\section{Proposed design flow}
The main innovation of the design flow proposed in \cite{ratto2023} is the possibility of supporting runtime adaptivity in the hardware accelerator through reconfiguration.

\subsection{Tools}
Different tools are utilized along the design flow:
\begin{itemize}
    \item the ONNXParser\footnote{https://gitlab.com/aloha.eu/onnxparser}, a Python application intended to parse the ONNX models and automatically create the code for a target device. It is composed of a Reader and many Writers, one for each target;
    \item The Vivado HLS tool\footnote{https://www.AMD.com/support/documentation-navigation/design-hubs/dh0012-vivado-high-level-synthesis-hub.html}, which synthesizes a C or C++ function into RTL code for implementation on AMD FPGAs. The resulting hardware can be optimized and customized through the insertion of directives in the code;
    \item The \ac{mdc}\footnote{https://mdc-suite.github.io/} is an open-source tool that can offer optional Coarse-Grained reconfigurability support for hardware acceleration. It takes as input the applications specified as dataflow, together with the library of the HDL files of the actors. These dataflows are then combined, and the resulting multi-dataflow topology is filled with the actors taken from the HDL library.
\end{itemize}

\subsection{Design Flow}\label{ss:df}
\begin{figure}[htbp]
\centerline{\includegraphics[trim=10cm 4cm 10cm 4cm, clip, width=.7\columnwidth]{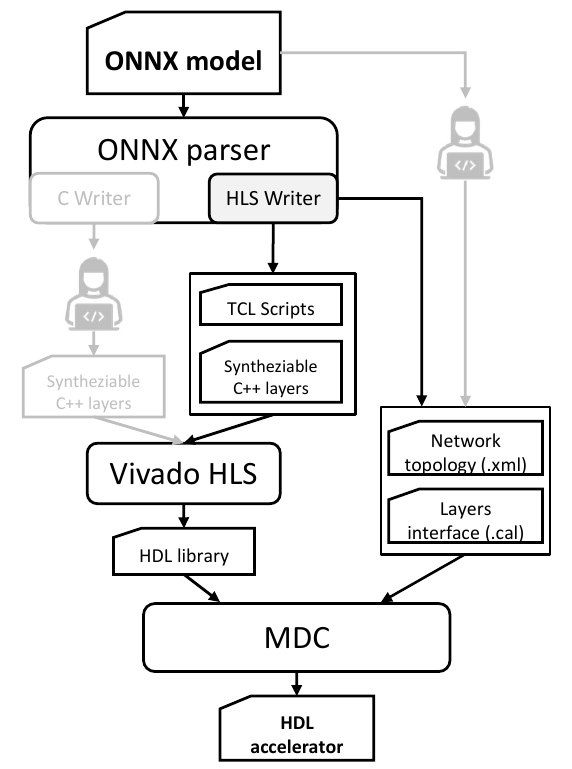}}
\caption{ONNX-to-Hardware design flow for the generation of adaptive neural-network accelerators on FPGAs. The manual steps needed in \cite{ratto2023} (grey lines) have been fully automated with the newly engineered \emph{HLS Writer}.}
\label{fig::design-flow}
\end{figure}

\begin{figure}[htbp]
\centerline{\includegraphics[trim=7cm 5.8cm 8cm 6cm, clip, width=.6\columnwidth]{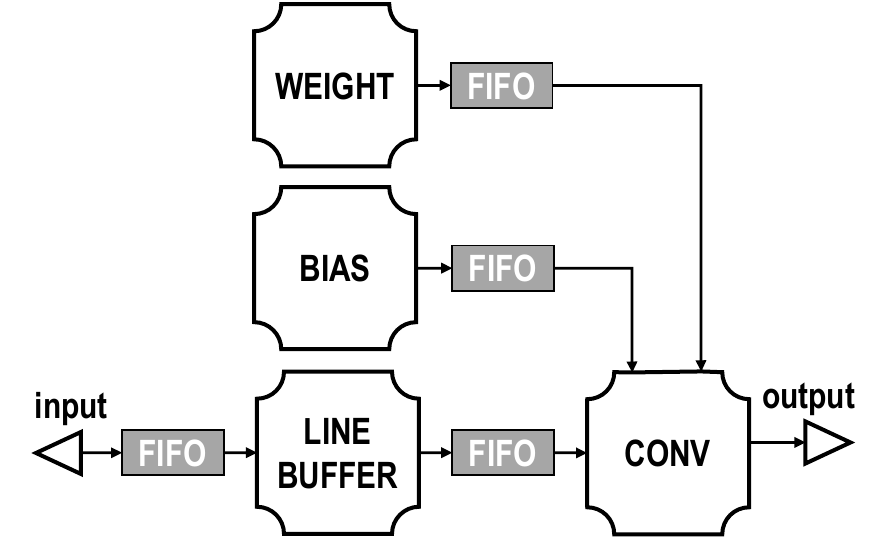}}
\caption{Representation of the streaming-based template architecture for a convolutional layer.}
\label{fig}
\end{figure}

The proposed flow, displayed in Figure~\ref{fig::design-flow}, starts from the ONNX representation of the \ac{nn} and produces a streaming accelerator that accelerates the input model. 
This file is given as input to the ONNX Parser: initially, the Reader reads the ONNX file and produces an intermediate format with a list of objects that describes layers and connections of the ONNX model. Then, the selected Writer creates the target-dependent files. When the target is the HLS flow, it is possible to customize the data precision used to represent weights and activations. 
The HLS Writer produces C++ files that implement the layers, and the TCL scripts to automate the synthesis by Vivado HLS. 
The C++ description of the layers is based on a template architecture: for the CONV layer, the core of the CNN, the template is composed of a Line Buffer actor that stores the input stream to provide data reuse; the Convolutional actor, whose function is to execute the actual computation; and the Weight and Bias actors that store the kernel parameters needed for the convolution. The resulting template is depicted in Figure~\ref{fig}. Each actor is developed to be customizable with the hyperparameter, e.g. input and kernel size, extracted from the ONNX model.
The HDL library produced by Vivado HLS is given as input to the Multi-Dataflow Composer, together with the XDF file that describes the topology of the network and the CAL files that identify the different actors. These latter are generated by the HLS Writer. Finally, the HDL file of the complete dataflow is automatically generated. 
Optionally, the \ac{mdc} Co-processor generator can be used to deploy the accelerator using the Vivado design suite. The Co-processor generator delivers the necessary scripts to wrap the accelerator  
and connect it to a complete processor-coprocessor system. Along with the hardware system, the drivers to call the coprocessor from the SDK application are made available.

In the work of Ratto et al. presented in \cite{ratto2023} the flow was semi-integrated and, as part of this preliminary study, the entire generation process from the ONNX file down to the accelerator deployment is fully automated.

\section{Preliminary results and future direction}

\begin{table*}[htb]
\renewcommand\arraystretch{1} 
\caption{Results of exploration with mixed precision data on an accelerator made of 2 convolutional blocks (consisting of a convolutional layer, max pooling,
batch normalization, and ReLU activation layers) followed by 1 fully connected layer. The accelerator classifies samples from the MNIST dataset. The model is quantized using post-training quantization. In the \emph{Datatype} column, Dx-Wy denotes that x bits are used to represent activations and y bits are used to represent parameters in fixed-point precision. The reported results target a Zynq7000, comprising a ``xc7z020-1csg484ces'' chip, and have been retrieved through post-synthesis simulations.}
	\centering
\begin{tabular}{l c c c c c c c c c c}
\toprule
\multirow{2}{*}{Datatype}     &Zero-weights            &LUT    &FF     &DSP    &BRAM  &Latency  &Throughput     &Power &Energy  &Accuracy\\
       &[\%]          &[\%] &[\%]  &[\%]  &[\%] &[us] &[FPS]      &[mW] &[uJ]  &[\%]\\ \midrule
D32-W32   &0.0   &29.6 &24.5 &29.5 &15.4 &1530 &88K &28.6 &43.7 &98\\
D16-W16   &0.0   &23.4 &20.2 &52.7 &15.4 &1510  &89K &25.3 &38.3 &98\\
D8-W16    &0.8   &9.1 &5.6 &15.5 &13.2 &510 &296K &20.1 &10.2 &76 \\
D16-W8    &15.0  &8.5 &0.6 &15.5 &13.2 &510  &296K &19.5 &9.9 &98\\
D16-W4    &55.3  &7.7 &4.3 &15.5 &4.3 &510 &296K &17.5 &8.9 &97\\
D16-W2    &85.7  &7.7 &4.3 &15.5 &4.3 &1140 &117K &15.0 &17.1 &68\\ \bottomrule
\end{tabular}
\label{table::results}
\end{table*}

To assess the re-engineered flow described in Section~\ref{ss:df}, a wide exploration targeting the MNIST classifier has been carried out, as described in Table~\ref{table::results}. The intent was also to show the impact of quantization on both model accuracy and hardware performance, which is generally in line with the expectations that \ac{ac} can offer, as discussed in Section \ref{ss:ac}.
It can be noticed that accuracy is not as affected by reducing parameter precision as it is by reducing activations precision. Moreover, reduced parameter precision leads to a reduced memory footprint (BRAM column) and a high percentage of zero weights. This latter can be exploited to combine quantization with pruning, which skips multiplications by zero. 
To have a fair comparison with state-of-the-art solutions presented in Table~\ref{table::comparison}, onboard-running experiments that consider also memory accesses are needed. However, we can see that the preliminary results show competitive performance in terms of utilized resources and latency/throughput. A broader comparison against state-of-the-art, based on significant onboard measurements and targeting more complex datasets, will be carried out in the future. Nonetheless, it is worth recalling that state-of-the-art approaches are not conceived to support runtime adaptivity, which is motivating our research instead.

Indeed, our future work intends to explore mixed precision in adaptive \ac{nn} accelerators. To have available a fully automated flow, with reconfiguration support capabilities, was a key preliminary step to save the manual effort in the accelerator definition and exploration. The analysis carried out so far on non-reconfigurable accelerators shows that promising trade-offs are present, e.g. trading off accuracy for reduced energy consumption. The ultimate goal will be the efficient runtime management of the system that implies, as a first step, the combination of the different working points over a reconfigurable substrate. This latter can certainly be achieved by leveraging on the whole set of functionality offered by the \ac{mdc} tool to design and operate reconfigurable and evolvable \ac{nn} accelerators for \ac{cps}, including the one presented in this study. Resulting accelerators will be able to switch configuration at runtime to adapt to the desired goal, e.g. when a limited energy budget is left a reduction in energy consumption is worth the cost of some accuracy loss.

One of the challenges we expect to face in this research's future steps is the limited onboard memory, which could constrain us to the execution of relatively small models (e.g. TinyML), especially when runtime switching among algorithms/configurations is required. The adoption of a reconfigurable approach, capable of sharing weights among configurations, should help us tackle that issue limiting the impact on the memory footprint of having more than one network available. This may limit the advantages in terms of accuracy achievable with Quantization Aware Training~\cite{gholami2021}. However, the preliminary results with Post Training Quantization show a limited drop in accuracy even with 4-bit weights.

\section*{Ackowledgments}
The authors would like to thank Stefano Esposito for his contribution to this work during his Master's thesis.

\end{document}